\documentclass[a4paper]{article}
\usepackage{amsmath}
\usepackage{amssymb}
\usepackage{graphicx}
\usepackage{float}
\usepackage{authblk}

\title{Radiotherapy Dosimetry: A Review on Open-Source Optimizer}
\author[1,2,3]{Paul Dubois}
\affil[1]{TheraPanacea}
\affil[2]{MICS, CentraleSupélec}
\affil[3]{Institut du Cancer de Montpellier}

\newcommand{\abs}[1]{|{#1}|}

\begin{document}
	
	\maketitle
	
	\begin{abstract}
		Radiotherapy dosimetry plays a crucial role in optimizing treatment plans for cancer patients.
		In this study, we investigate the performance of a dozen standard state-of-the-art open-source optimizers for radiotherapy dosimetry.
		Our evaluation includes the use of TGG119 benchmark cases as well as one real case obtained from the Institute du Cancer de Montpellier (ICM).
		Among the tested optimizers, Newton CG demonstrates the fastest convergence in terms of the number of iterations.
		However, when considering the computation time per iteration, LBFGS emerges as the most efficient optimizer.
		These findings shed light on the performance of open-source optimizers for radiotherapy dosimetry, aiding practitioners in selecting suitable optimization tools for efficient treatment planning.
	\end{abstract}
	
	\section{Introduction}
	Radiotherapy, a widely utilized intervention for cancer treatment, employs ionizing radiation to eliminate malignant cells.
	Intensity-modulated radiation therapy (IMRT) has emerged as a notable technique within radiotherapy, aiming to deliver high radiation doses to tumors while minimizing exposure to healthy surrounding tissues \cite{ReportIMRT2003}.
	Traditional IMRT strategies typically employ a set number of beams, often 5, 7, or 9, originating from various angles around the patient, commonly distributed evenly \cite{Bortfeld_2006}.
	Each beam's intensity is modulated to optimize the delivery of radiation doses to the tumor while reducing exposure to healthy tissues.
	This approach surpasses the effectiveness of the 3D-conformal radiotherapy (3D-CRT) technique \cite{KOLE2012} \cite{PALMA2008} \cite{VERHEY1999}.
	To facilitate precise and efficient radiation delivery, a computer-controlled device called the multi-leaf collimator (MLC) is utilized to shape the radiation beam according to the contours of the tumor.
	
	The effectiveness of a radiotherapy treatment plan relies on the optimization procedure, which involves a series of steps aimed at ensuring the optimal delivery of radiation in accordance with the prescribed guidelines of medical practitioners.
	Typically, computer software is employed to facilitate the optimization process, taking into account various factors such as patient anatomy, the size and location of the tumor and organs, and the radiation objectives defined by medical professionals.
	
	While there has been comparison between commercial software\cite{Tang2019} \cite{Mora-Ramirez2020}, the aim of this paper is to investigate the best optimizer for this task among the open source optimizers.
	
	\paragraph{Pre-dose-optimization}
	The initial stage of the optimization process entails the creation of a virtual representation of the patient's anatomical structure using advanced medical imaging modalities, such as computed tomography (CT) or magnetic resonance imaging (MRI) scans.
	This model is subsequently utilized to accurately determine the size and location of the tumor, as well as to delineate the surrounding healthy tissues that necessitate protection from radiation exposure.
	Following this, the radiation dose required for effective treatment is established, typically based on dose-volume objectives defined by physicians (e.g., ensuring that 95\% of the planning target volume receives a minimum dose of 75 Gy).
	Determining the appropriate dose takes into consideration factors such as tumor characteristics, location, size, as well as the patient's medical history and overall health status.
	These essential steps in the optimization process are carried out by medical professionals with expertise in radiotherapy treatment planning.
	
	\paragraph{Radiotherapy doses}
	The subsequent step entails the computation of the radiation dose distribution within the patient's volumetric anatomy.
	This is achieved by simulating a particular configuration of the multi-leaf collimator (MLC) on the patient's body, utilizing the available medical imaging data.
	The resulting computed dose represents a mapping from the three-dimensional volume of the patient's anatomy to a scalar value measured in Grays (Gy), which denotes the absorbed radiation energy.
	In practical implementation, a discrete representation of the dose distribution is utilized, wherein the dose is calculated for each individual voxel comprising the patient's anatomical structure.
	
	\paragraph{Dose-Volume Histograms}
	Medical professionals have meticulously identified and delineated the pertinent anatomical structures within the patient's anatomy.
	This allows the computation of dose-volume histograms (DVHs) for each structure, predicated on a specified dose distribution.
	The dose-volume objectives are subsequently represented as specific points on the DVH curve, which correspond to the desired minimum or maximum dose constraints.
	These objectives delineate the desired thresholds that should be upheld, with points on the DVH curve either located above (for minimum dose constraints) or below (for maximum dose constraints) the prescribed thresholds.
	
	\paragraph{Dose Evaluation}
	Physicians employ multiple criteria to assess the quality of a radiation dose administered during treatment.
	Initially, they scrutinize the three-dimensional distribution of the dose across the patient's anatomy, focusing on the spatial allocation among different anatomical structures, as well as identifying the presence, number, and locations of regions with excessive radiation (referred to as "hot spots").
	Subsequently, physicians conduct a thorough analysis of the dose-volume histograms (DVHs) to evaluate the degree of compliance with predefined DVH objectives. This crucial evaluation step aims to safeguard the adjacent healthy tissues from unnecessary radiation exposure.
	By optimizing the treatment plan and meticulously assessing the quality of the dose distribution, physicians strive to ensure the attainment of the most favorable outcome for the patient.

	\section{Methods}
	To ensure the precision and effectiveness of radiation therapy, a robust dose optimization process are essential.
	
	\subsection{Radiotherapy Dosimetry}
	Radiotherapy dose optimization can be conceptualized as an inverse problem, whereby the objective is to determine the most suitable radiation dose distribution that aligns with the desired treatment outcome \cite{Webb2003}.
	In other terms, the challenge lies in identifying the radiation intensity or fluence maps that deliver the prescribed dose to the tumor while minimizing exposure to healthy tissues.
	
	Mathematical optimization algorithms are employed to address the inverse problem of radiotherapy dose optimization.
	These algorithms aim to identify the optimal solution by minimizing a predefined objective function that encompasses treatment goals and constraints.
	Typically, the objective function includes terms that penalize both underdosing and overdosing of the tumor and overdosing of surrounding healthy tissues.
	It may also incorporate terms that account for the complexity or deliverability of the treatment plan.
	
	Efficiently solving this optimization problem often involves designing the objective function to be convex, thereby providing a well-defined target for the optimization process.
	Gradient-based methods, Newtonian algorithms, or quasi-Newtonian algorithms are commonly employed for this purpose.
	We aim at benchmarking state-of-the-art open-source optimization algorithms for the specific task of radiotherapy dosimetry.
	
	\subsection{Data}
	In this research endeavor, our focus was to evaluate the various open-source optimizers. 
	We used the widely recognized TG-119 \cite{AAPM-TG119} cases as a benchmark for evaluating radiation therapy plans optimization.
	The TG-119 dataset provides specific dose goals, which we incorporated into our proposed cost function.
	
	We also used one real case of prostate cancer treatment from ICM.
	For this case, doctors had provided specific dose goals, that we again incorporated into our proposed cost function.
	
	The TGG 119 multiple PTVs is a theoretical case, unlikely to happen in real life.
	However, the three other cases represent a comprehensive set of what dosimetrists could encounter on a daily basis.
	
	The simulation of the beams was done using TheraPanacea dose engine, which uses collapse cone convolution techniques, and is conformal to other simulator available on the market.
	
	\subsection{Objective function}
	The cost function is formulated as a weighted sum of multiple objectives, with each objective corresponding to a specific dose goal. The formulation is as follows:
	
	\[f(\mathbf{d}) = \sum_{o \in \mathcal{O}} w_o f_o(\mathbf{d})\]
	
	where:
	\begin{itemize}
		\item \(\mathbf{d}\) represents the dose distribution at the voxel level, and \(\mathbf{d}[s]\) denotes the dose on voxels within the structure \(s\)
		\item \(\mathcal{O}\) denotes the set of objectives corresponding to dose volume goals
		\item \(w_o\) signifies the weight assigned to the objective \(o \in \mathcal{O}\)
		\item \(o_s\), \(o_d\), and \(o_v\) refer to the structure, dose, and volume goals of the objective \(o \in \mathcal{O}\)
	\end{itemize}
	
	The objective function \(f_o(\mathbf{d})\) is computed based on the specific type of dose volume constraint:
	If \(o\) represents a maximum dose volume constraint\footnote{e.g.: top 20\% of the volume should receive at most 30 Gy}, \(f_o(\mathbf{d})\) is calculated as \(\sum_{d \in \mathbf{d}[o_s]}(d - o_d)_+^2\); if \(o\) represents a minimum dose volume constraint\footnote{e.g.: 95\% of the volume should receive at least 70 Gy}, then \(f_o(\mathbf{d})\) is calculated as \(\sum_{d \in \mathbf{d}[o_s]}(o_d - d)_+^2\).
	The formulation involves a squared over/under-dose penalty function.
	
	In addition to the above, we introduced a regularization term that penalizes variations in bixel values between neighboring regions, also employing a squared penalty.
	
	The optimization process involves finding the optimal bixel values (\(\mathbf{b}\)) by solving \(\mathbf{d} = \mathbf{L}\mathbf{b}\), where \(\mathbf{L}\) is a precomputed dose-influence matrix mapping bixels to voxels.
	Notably, since negative energy rays are physically infeasible, we ensured that each bixel value is non-negative (\(b \geq 0 \quad \forall b \in \mathbf{b}\)).
	To achieve this, we computed \(\mathbf{d} = \mathbf{L}\abs{\mathbf{b}}\), where \(\abs{\mathbf{b}}\) denotes the element-wise absolute value of \(\mathbf{b}\).
	
	By construction, the objective function is convex.
	Consequently, minimizing the objective function with a given set of weights should invariably converge to the same radiotherapy plan.
	To generate different treatment doses for the same patient case, dosimetrists can play withe the weights of each sub-objective function.
	This is outside the scope of this small review article, so we decided to just set the weights of all constraints equal to one.
	
	\subsection{Open-source Optimizers}
	We tried to have a comprehensive test of available open-source optimizers, here is a short description of the ones tested:
	
	\paragraph{(Stochastic) Gradient Descent}
	Is an optimization algorithm that iteratively updates the model parameters in the direction of the negative gradient of the objective function.
	In our case, it is not stochastic, since it calculates the gradient using the current solution\footnote{Our objective function has all its inputs as parameters, so there is no notion of stochasticity.} \cite{Lemarechal2012}.

	\paragraph{Conjugate Gradient}
	Is an iterative optimization algorithm commonly used to solve systems of linear equations or quadratic optimization problems.
	It iteratively computes conjugate directions and updates the solution along these directions, aiming to minimize the objective function \cite{Hestenes1952}.
	Conjugate Gradient is often applied in scenarios where the Hessian matrix is unavailable or expensive to compute.

	\paragraph{Newton}
	Newton's method is an iterative optimization algorithm that uses the second-order derivative (Hessian matrix) to find the minimum of a function.
	It updates the current estimate by taking into account both the first-order derivative (gradient) and the second-order derivative \cite{nocedal06}.
	
	\paragraph{SLSQP}
	(Sequential Least Squares Programming) is a sequential quadratic programming algorithm used for constrained optimization.
	It iteratively solves a sequence of quadratic programming subproblems to find the optimal solution subject to constraints \cite{Bonnans2006}.
	
	\paragraph{RMSprop}
	(Root Mean Square Propagation) is an optimization algorithm that addresses the problem of diminishing learning rates in traditional gradient descent methods.
	It divides the learning rate by the root mean square of the past gradients, which helps to stabilize and speed up convergence \cite{hinton2012}.
	
	\paragraph{BFGS-based}
	\subparagraph{Pure BFGS}
	(Broyden-Fletcher-Goldfarb-Shanno) is a quasi-Newton method that approximates the Hessian matrix using updates based on gradient information.
	It performs a line search to determine the step size that minimizes the objective function along the search direction \cite{fletcher1987}.
	\subparagraph{L-BFGS}
	(Limited-memory BFGS) is a variation of BFGS that uses a limited-memory approach to approximate the Hessian matrix.
	It stores a limited number of past gradient and parameter values to compute an approximate inverse Hessian matrix efficiently \cite{Liu1989}.
	
	\paragraph{Adam-based}
	\subparagraph{Pure Adam}
	(Adaptive Moment Estimation) is an optimization algorithm that combines ideas from both adaptive learning rates and momentum methods.
	It computes adaptive learning rates for each parameter based on estimates of the first and second moments of the gradients \cite{kingma2017}.
	\subparagraph{RAdam}
	(Rectified Adam) is a variant of the Adam optimizer that introduces a rectification term to stabilize the adaptive learning rate.
	It aims to address some convergence issues that can occur in Adam by dynamically adjusting the variance of the adaptive learning rate \cite{liu2021}.
	\subparagraph{NAdam}
	(Nesterov Adam) combines the Nesterov accelerated gradient method with the Adam optimizer.
	It incorporates Nesterov momentum into the Adam update rule to improve convergence and provide better generalization \cite{tato2018}.
	\subparagraph{AdamDelta}
	Is another variant of the Adam optimizer that replaces the second moment estimates (variance) with a delta parameter.
	It eliminates the need for storing and updating the moving average of the squared gradients, which can be beneficial in memory-constrained settings \cite{zeiler2012}.
	\subparagraph{Adamax}
	Is an extension of the Adam optimizer that uses the infinity norm (max norm) of the gradients instead of the L2 norm. It is designed to handle sparse gradients more effectively and can be particularly useful in deep learning models \cite{bera2020}.

	\paragraph{Rprop}
	(Resilient Backpropagation) is an optimization algorithm specifically designed for neural networks.
	It updates the weights based on the sign of the gradient, adjusting the step size adaptively.
	Rprop performs weight updates independently for each weight parameter \cite{Riedmiller1992}.
	
	\paragraph{Other optimizers variations}
	In addition, we also tested AdamW, Adagrad and ASGD.
	However, AdamW \& Adagrad behaved similarly to Adam, and ASGD behaved similarly to SGD.
	For readability purposes, we did not include them in the results.

	\section{Results}
	
	\begin{figure}
		\centering
		\includegraphics[width=0.95\textwidth]{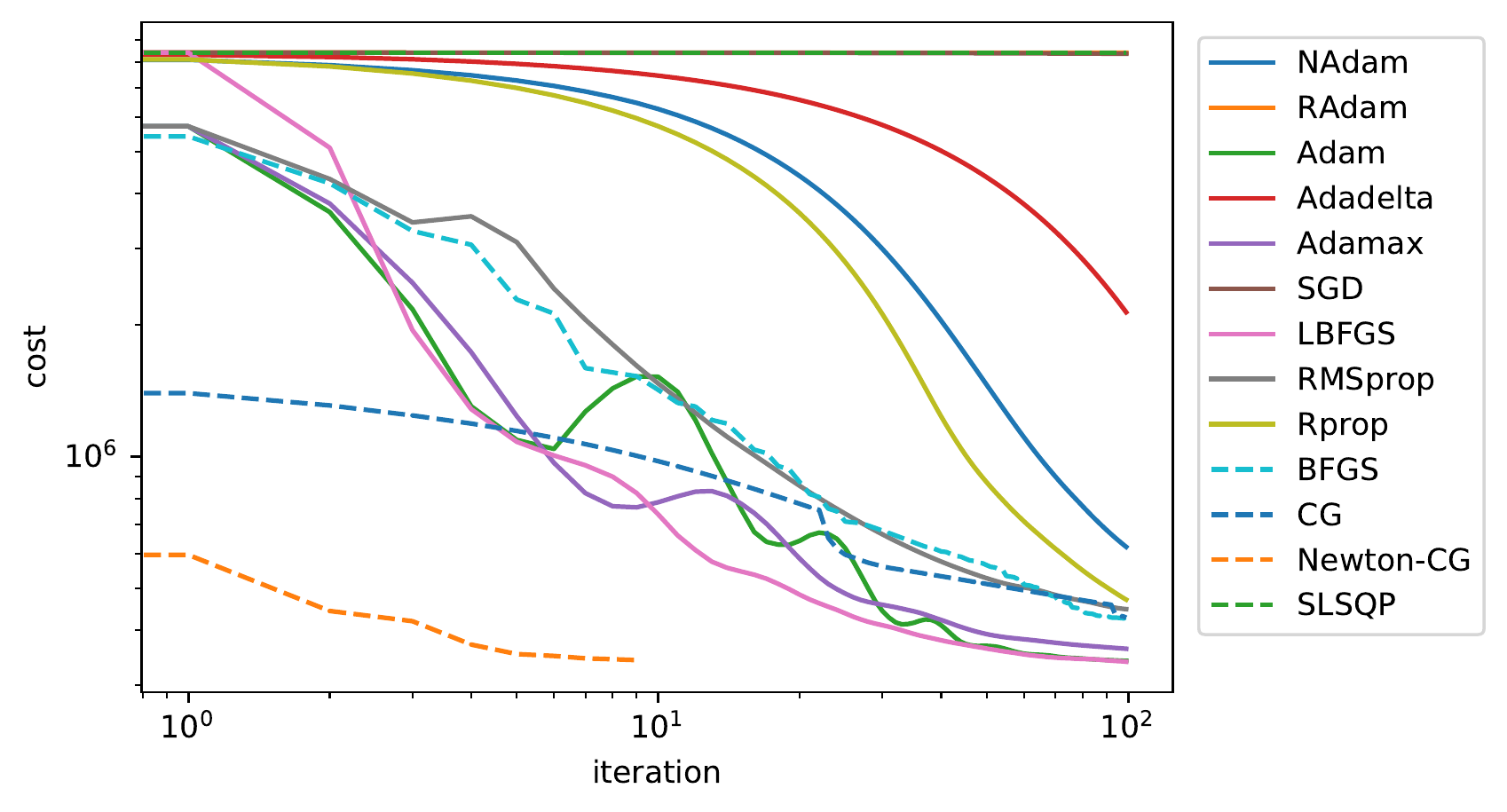}		\includegraphics[width=0.95\textwidth]{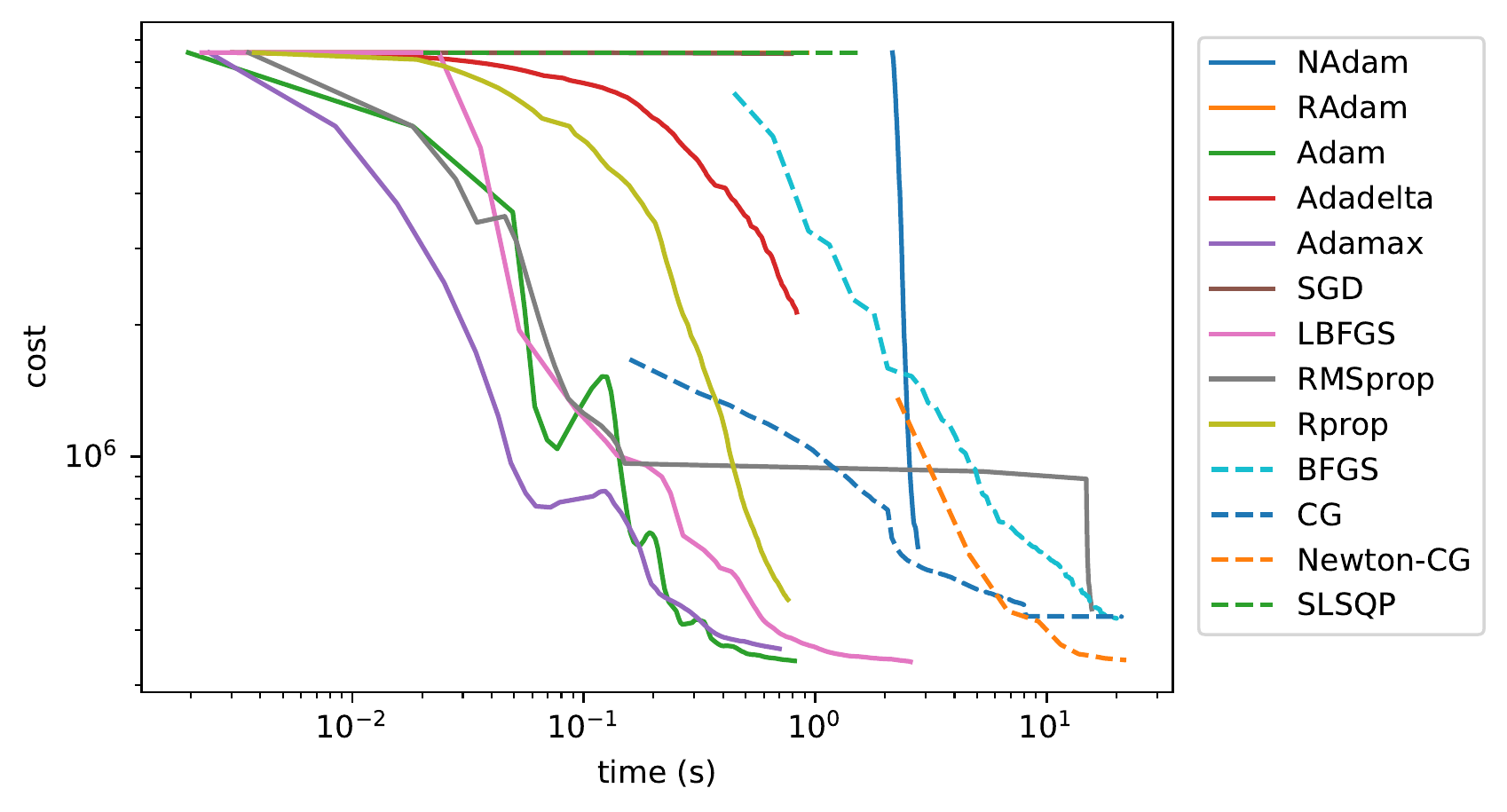}
		\caption{TGG 119: Multiple PTVs}
		\label{fig:tgg119multi}
	\end{figure}
	
	\begin{figure}
		\centering
		\includegraphics[width=0.95\textwidth]{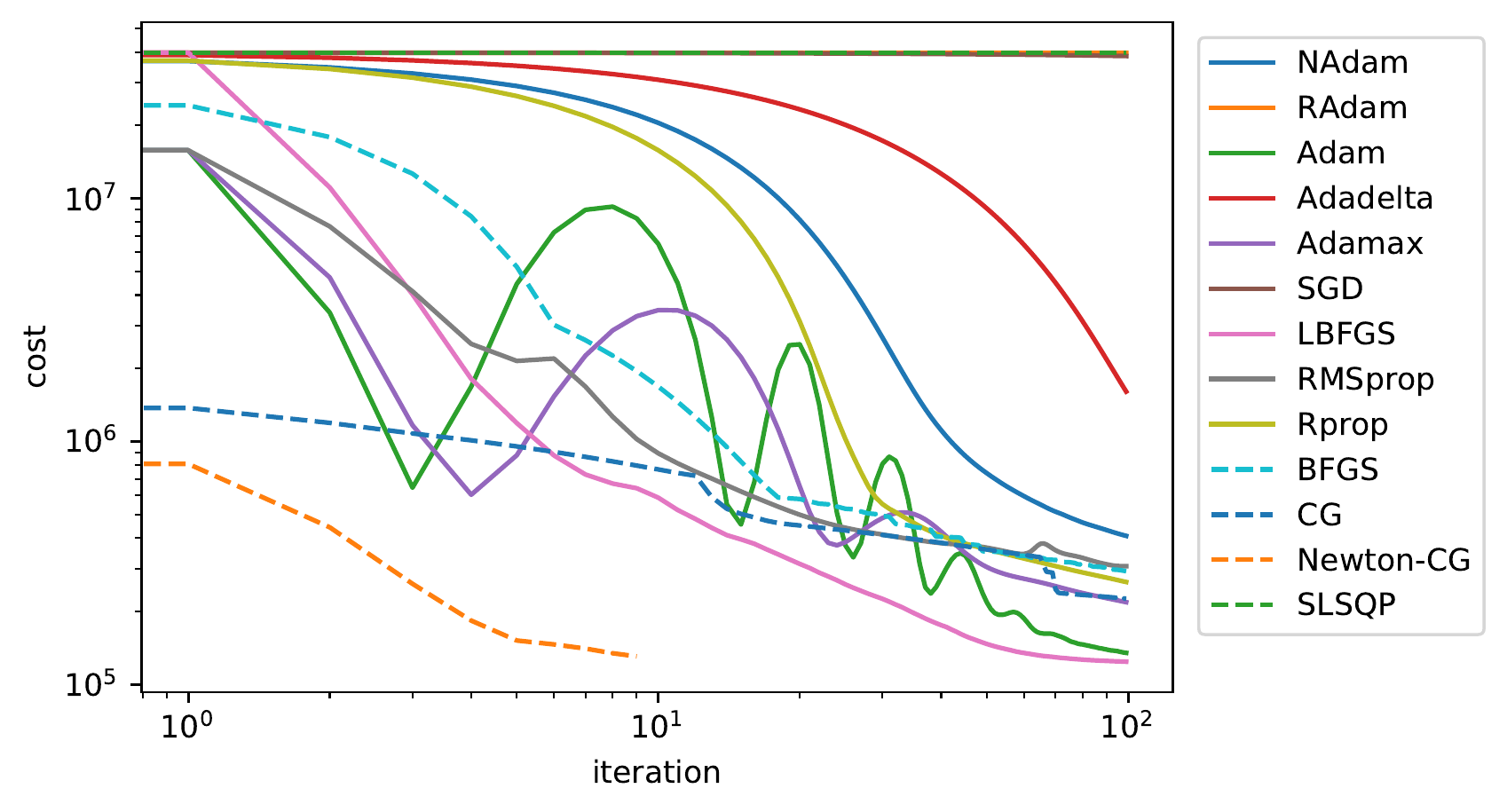}		\includegraphics[width=0.95\textwidth]{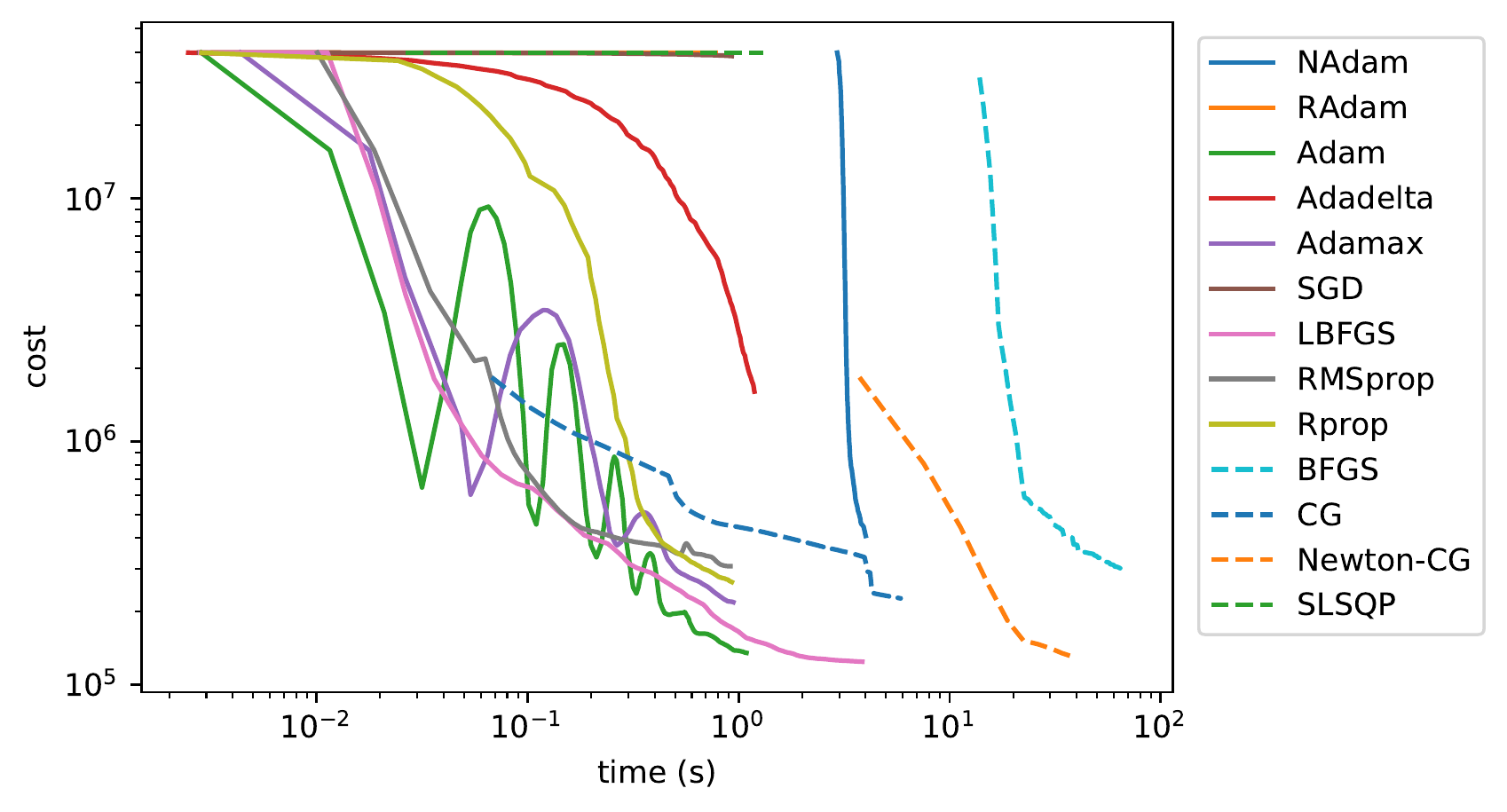}
		\caption{TGG 119: Head and Neck}
		\label{fig:tgg119hn}
	\end{figure}
	
	\begin{figure}
		\centering
		\includegraphics[width=0.95\textwidth]{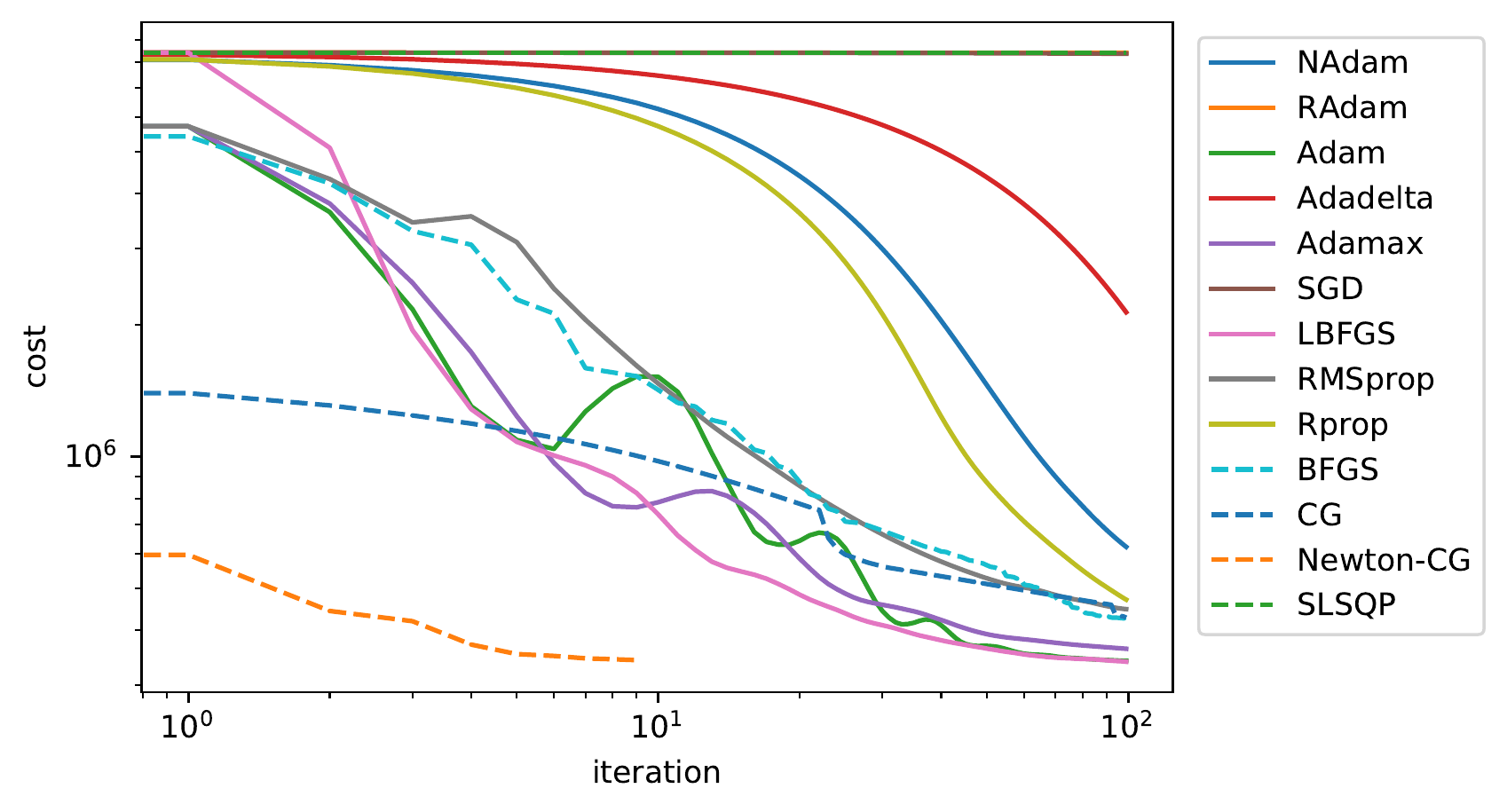}		\includegraphics[width=0.95\textwidth]{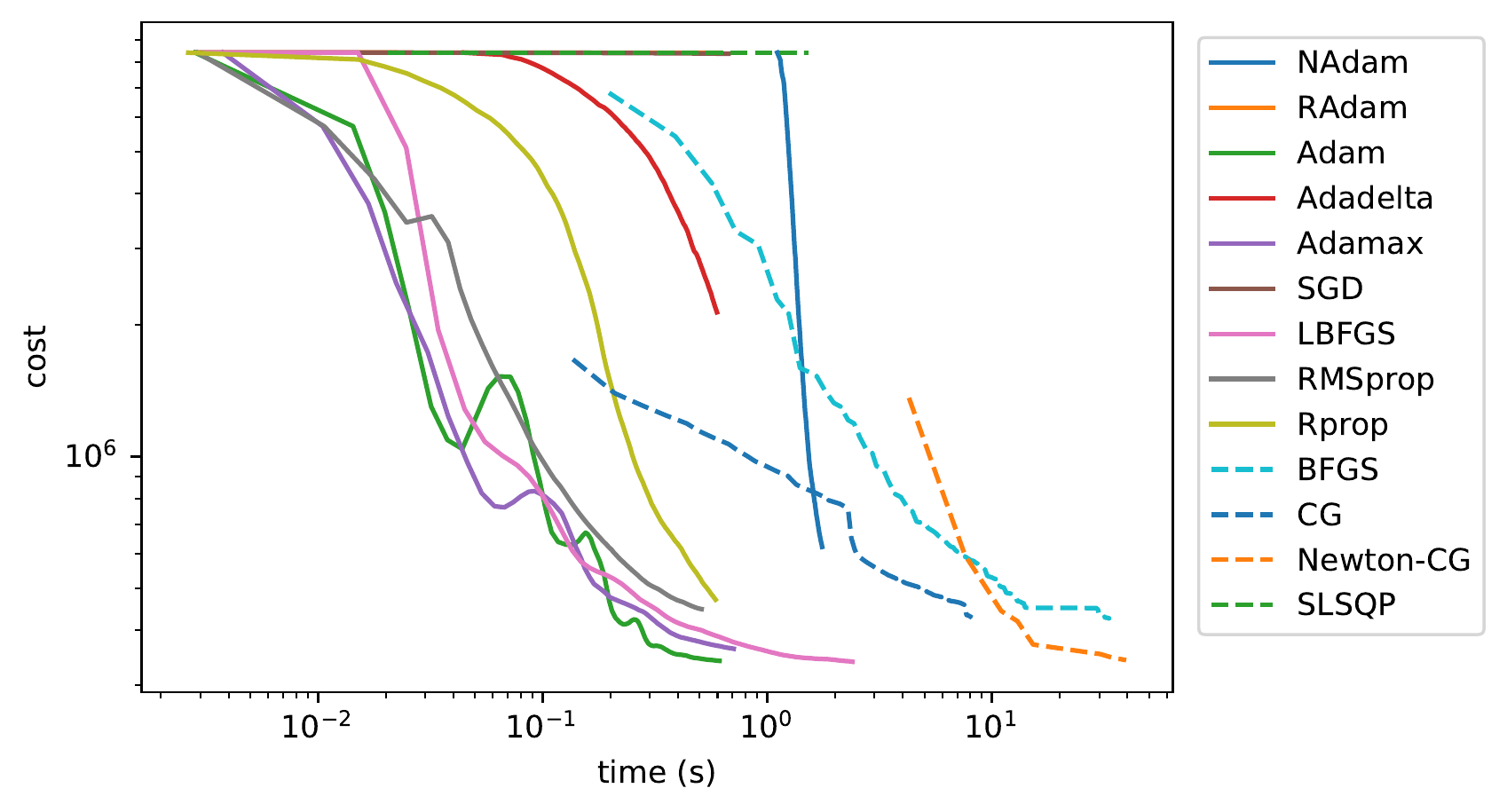}
		\caption{TGG 119: Prostate}
		\label{fig:tgg119prostate}
	\end{figure}
	
	\begin{figure}
		\centering
		\includegraphics[width=0.95\textwidth]{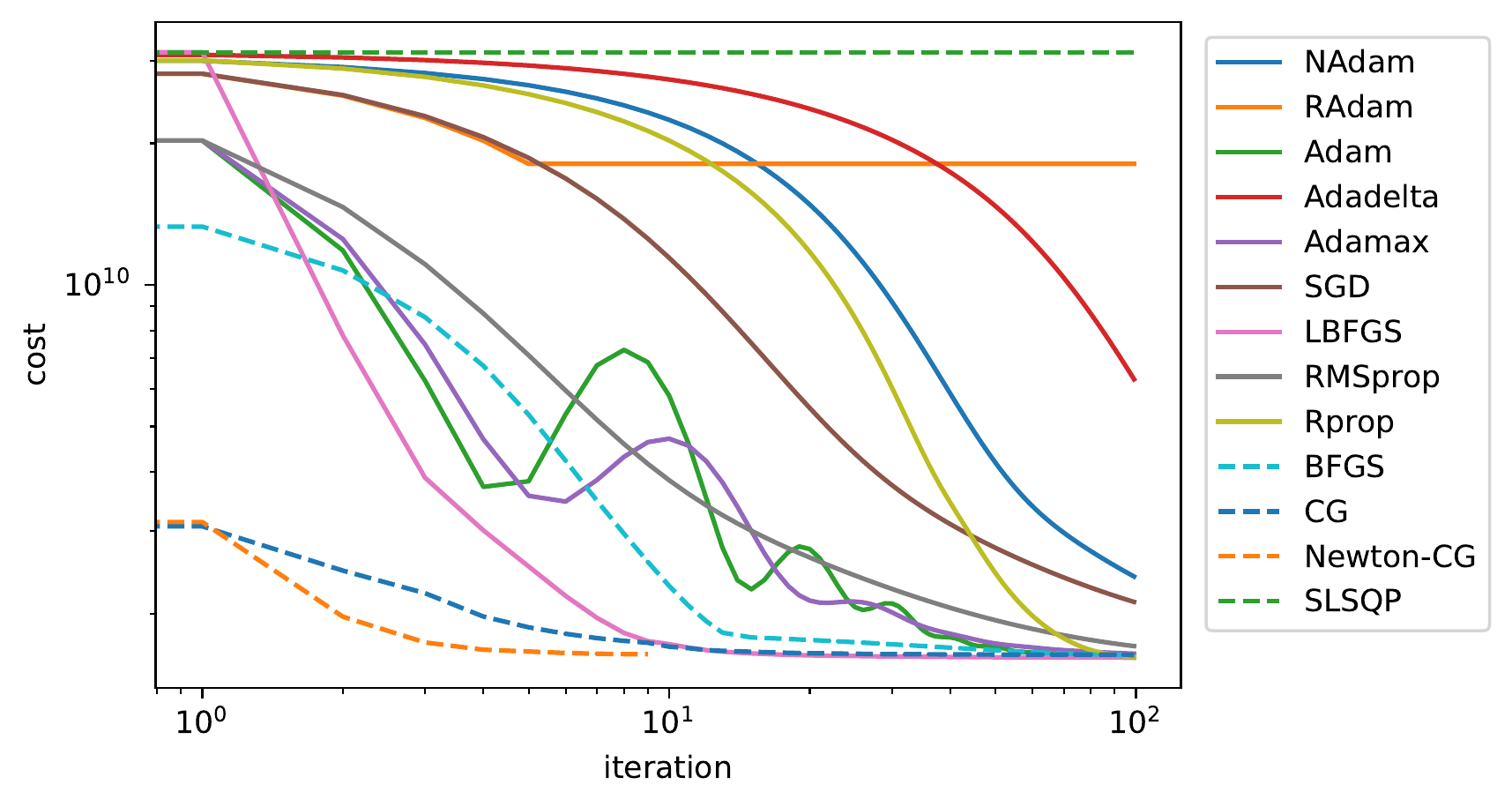}		\includegraphics[width=0.95\textwidth]{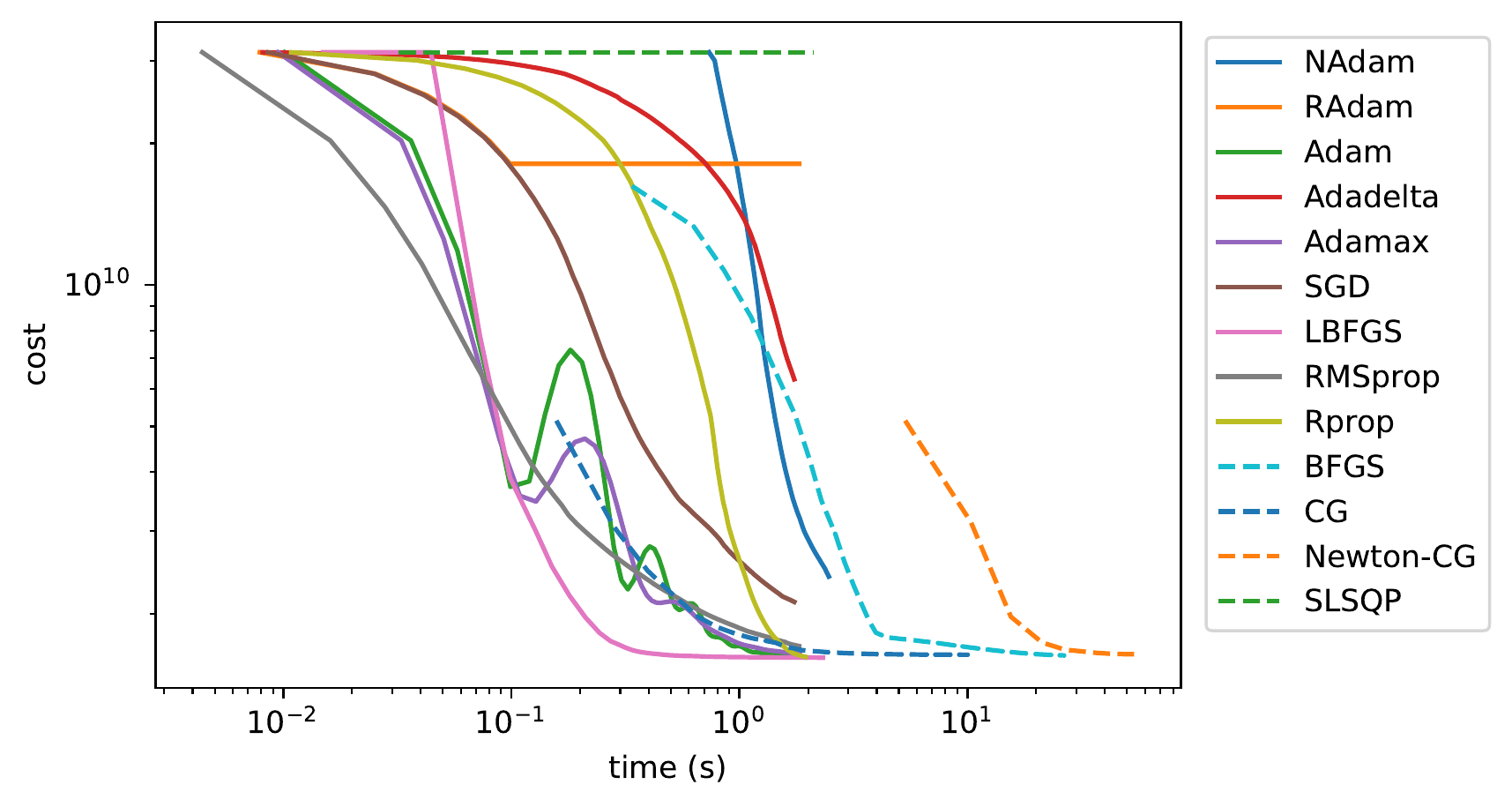}
		\caption{ICM: (Typical) Prostate}
		\label{fig:ICMprostate}
	\end{figure}
	
	\paragraph{Newton's method}
	Based on the iterations-wise graph analysis, Newton's method emerges as the most optimal, consistently achieving a stable converged state within a mere 10 steps across all four examined cases.
	However, Newton's method steps are very expensive to compute, since it uses second order derivative (the Hessian) that is expensive to compute (in terms of calculation time).
	
	It is widely recognized that Newton's method excels in optimizing convex functions \cite{PoczosTibshirani2013}.
	Our objective function is built to be convex.
	Hence, it make sense that this optimization algorithm is particularly effective.
	
	\paragraph{Best Algorithms}
	Other than Newton's method, three algorithms have similar performances: Adam, RMSprop and LBFGS.
	Adam (and RMSprop, to a lesser extent) appear to have more "wavy" cost curves, while LBFGS cost decreases in a more stable way.
	These observations are true both iteration and time-wise.
	
	TGG 119 Multiple PTVs (figure \ref{fig:tgg119multi}) is the smallest problem, and th real ICM prostate case (fig. \ref{fig:ICMprostate}) is the largest problem (in terms of patient/organs/structure volume size); TGG 119 fake head \& neck (fig. \ref{fig:tgg119hn}) and TGG 119 fake prostate (fig. \ref{fig:tgg119prostate}) have similar sizes.
	Notably, there is an observable trend indicating that as the problem size increases, LBFGS outperforms both RMSprop and Adam optimization algorithms.
	
	\paragraph{LBFGS vs BFGS}
	It would be expected that BFGS performs better than LBFGS in terms of iterations, but not in terms of time (since LBFGS is a fast approximation of the BFGS technique).
	However, we observe that LBFGS outperforms BFGS even on the iterations-wise graph.
	This suggests that the limited memory approximation made are biased towards suitable directions in these type of problems.
	
	\section{Discussion}
	If it was possible to make Newton's method faster, than we would advise to use Newton's optimization algorithm.
	However, to the best of our knowledge, computing the Hessian remains long, not only in our implementation.
	
	Hence, we advise to use the LBFGS algorithm for the problem of dose optimization i radiotherapy; it appears to be the fastest to converge, and converged steadily on the four cases tested.

	\bibliographystyle{plain}
	\bibliography{refs}
\end{document}